\documentclass[a4paper]{jpconf}
\usepackage{graphicx}
\usepackage{dcolumn}

\usepackage{xcolor}
\usepackage{caption}
\usepackage{subcaption}
\usepackage{amsmath}
\usepackage{amsfonts}
\usepackage{mathtools}

\DeclareUnicodeCharacter{2212}{-} 
\usepackage[bookmarks=false]{hyperref} 

\begin{document}
\title{The velocity dependence of dry sliding  friction at the nano-scale}

\author{Rasoul Kheiri$^1$, Alexey A Tsukanov$^{1,2}$, and Nikolai V Brilliantov$^{1}$}

\address{$^1$ Skolkovo Institute of Science and Technology, Moscow, 121205, Russia}

\address{$^2$ Research and Development Centre, TerraVox Global Ltd., Paphos, 8310, Cyprus}

\ead{Rasoul.Kheiri@skoltech.ru}

\begin{abstract}
We performed molecular dynamics (MD) experiments to explore dry sliding friction at the nanoscale. We used the setup comprised of a spherical particle built up of 32,000 aluminium atoms, resting on a semi-space with a free surface, modelled by a stack of merged graphene layers. We utilized  LAMMPS with the  COMB3 many-body potentials for the inter-atomic interactions and Langevin thermostat which kept   the system at $300 K$. We varied the normal load on the particle and applied  different tangential force, which caused the particle sliding. Based on the simulation data, we demonstrate that the friction force $F_{\rm fr}$ linearly depends on the sliding velocity $v$, that is, $F_{\rm fr}=-\gamma v$, where $\gamma$ is the friction coefficient. The observed dependence is in a sharp contrast with the macroscopic Amontons-Coulomb laws, which predict the velocity independence of sliding friction. We explain such a dependence by  surface fluctuations of the thermal origin, which give rise to surface corrugation hindering  sliding motion. This mechanism is similar to that of the viscous friction force exerted on a body moving in  viscous fluid.  
\end{abstract}

\section{\label{sec:level1}Introduction}
According to  Amontons-Coulomb laws of friction, for a macroscopic solid object moving on a surface, dry sliding friction neither depends on the contact are of the bodies, nor on the sliding velocity. Instead, it is proportional to the normal load, e.g.  \cite{popov2010contact,berman1998amontons}.  In contrast, the viscous friction force, which emerges when a body moves through viscous fluid, do depends on the body's velocity. The form of the dependence may vary, but normally,  it increases with the increasing speed. Often a linear dependence of the friction force $F_{\rm fr}$ on the velocity $v$ is observed, $F_{\rm fr}= -\gamma v$, where $\gamma$ is the friction coefficient.  Such an important difference stems from  different mechanisms which give rise to the friction force.  In the former case, dry friction originates due to mesoscopic asperities on the surface of macroscopic bodies; the asperities engage with each other and hinder the motion. In the latter case, viscous friction emerges due to thermal fluctuation of the local molecular stress and hence it is determined by the temperature of the system. Obviously the mechanism of dry friction for macroscopic bodies is athermal, that is, it is not related to molecular fluctuations \cite{BrilPhylProc}.  

Molecular fluctuations may become important, however, at the meso and nanoscale, providing a significant impact on the dry friction. At  the mesoscale, studies of the static and dynamic friction and the transition to sliding demonstrated the importance of elastic deformations, see e.g.  \cite{rubinstein2004detachment,rubinstein2006contact,rubinstein2007dynamics,ben2010dynamics,ben2011static,lorenz2012origin}; moreover, it has been shown that the onset of sliding may vary  \cite{ben2011static}. In Ref. \cite{ben2010dynamics} deviations from the Amontons-Coulomb laws have been also reported --  the dependence of friction force on the contact area, and of friction coefficient on the normal load.

Similarly, recent studies in nanotribology \cite{urbakh2004nonlinear,mo2009friction,vanossi2013colloquium} demonstrated that solid friction at the nanoscale  can noticeably deviate from  that predicted  by  Amontons-Coulomb  macroscopic laws. For instance, in the absence of a net normal load in thin films, friction could be more like the dissipation force in liquids \cite{krim2012friction}, that is, it can presumable be  velocity dependent.  

The experimental investigation of dry solid friction at the nanoscale started, essentially, with the  invention of fine experimental tools like atomic force microscopy (AFM) \cite{binnig1986atomic,carpick1997scratching}, surface force apparatus (SFA) \cite{binnig1986atomic}, and quartz crystal microbalance (QCM) \cite{krim1988damping,krim1991nanotribology,krim2007qcm} (see also  \cite{persson2013slidng,mate2019tribology}). Still the most detailed view can be provided by computer experiments -- the numerical investigation on the atomic scale by means of molecular dynamics (MD) simulation \cite{frenkel2001understanding}. Since the number of particles simulated in MD is not "macroscopically" large, one needs to link the molecular system to a somewhat artificial system -- thermostat, which keeps the desired conditions \cite{frenkel2001understanding}. The most popular are  Anderson \cite{andersen1980molecular}, Nose-Hoover \cite{nose1984unified,nose1984molecular,hoover1985canonical,hoover1986constant}, and Langevin \cite{schneider1978molecular} thermostats.  Each of them have some advantages and disadvantages \cite{frenkel2001understanding}; sometimes thermostating may result in non-physical effects on the interface, so that further correction are to be applied in the simulations \cite{vanossi2013colloquium,benassi2012optimal,kantorovich2008generalized}. Hence  reliable data may be obtained only under a careful design of the model \cite{dong2013molecular}. 

In the present study we analyse dry sliding friction at the nanoscale by means of MD. We investigate the friction force acting on a spherical nanoparticle sliding on a solid surface - the boundary of the semi-space, which is modelled by a several layers of graphene stacked on each other. Molecular fluctuations of the thermal origin of the free surface yield surface corrugation. This hinders the motion of the particle on the surface and thus gives rise to the friction force. Since such a mechanism, associated with the thermal fluctuations is similar to the friction mechanism of the viscous friction, one expects that the friction force would be proportional to the sliding velocity $v$, that is $F_{\rm fr} = -\gamma v $. Where the friction coefficient $\gamma$ may depend on the normal load $F_{\rm N}$. 

The rest of the article is organized as follows in the next Sec. II we present the detail of the MD simulations and the way of obtaining the friction coefficient from the rough data. In Sec. III we discuss the MD results, with the focus on the dependence of the friction force on the sliding velocity. Finally, in Sec. IV we summarize our findings. 

 \section{Molecular dynamics simulations \label{MDs}}

\subsection{Simulation setup}
\begin{figure}[h]
\centering
\includegraphics[width=0.5\textwidth]{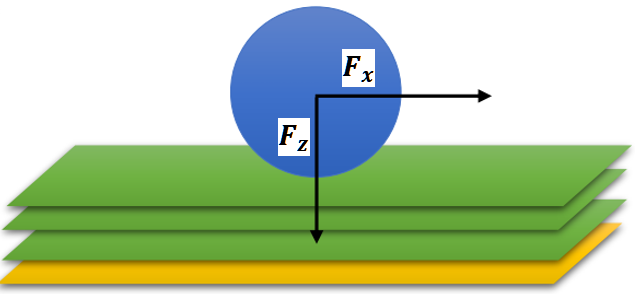}
\caption{Sketch of the simulation setup. The tangential, $F_x$,  and normal force, $F_z$, are exerted on the spherical aluminum particle built up of 32,000  atoms (blue). The semi-space with a free surface is modelled by a stack of graphene layers (green and yellow), with the fixed bottom layer (yellow). }
\label{sys2}
\end{figure}

\begin{figure}
\centering
\begin{subfigure}{0.49\textwidth}
    \includegraphics[width=0.99\textwidth]{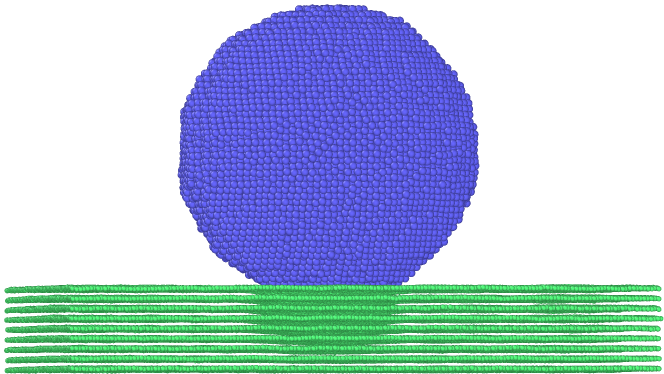}
    \caption{At the onset of simulation}
    \label{sys1:first}
\end{subfigure}
\begin{subfigure}{0.49\textwidth}
    \includegraphics[width=0.89\textwidth]{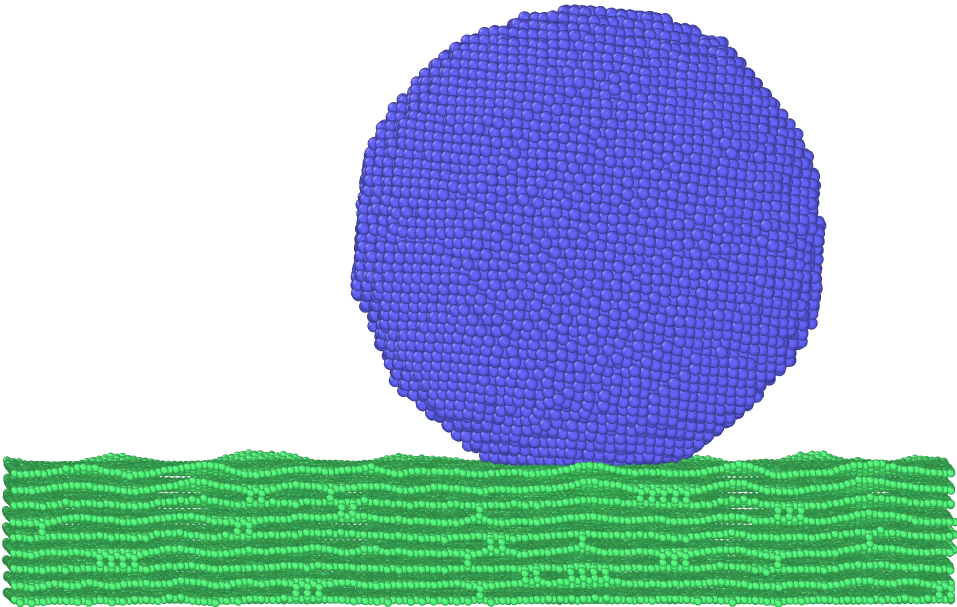}
    \caption{After some time steps.}
    \label{sys1:second}
\end{subfigure}    
\caption{Snapshots  of the simulated system -- aluminium spherical particle (blue) is  moving on several graphene layers (green) to the right. The exerted normal load is $F_z = 33.28 \times 10^{-9} N$, and the tangential force is $F_x = 4.1 \times 10^{-9} N$. The temperature is kept  around $T = 300 K$; the Langvian thermostat is applied.}
\label{sys1}
\end{figure}

MD simulations for a spherical particle moving on a fluctuating surface have been performed using LAMMPS \cite{thompson2016lammps}, see Figs.~\ref{sys2}-\ref{sys1} which provide the sketch of the model and simulation images.  The spherical particle has been made up of 32,000 aluminium atoms, interacting with a classical many-body potential (see the discussion below). To prepare the aluminum particle we use a cubic FCC aluminum lattice and then cut from it a sphere.    

The semi-space with the free surface has been constructed using a stack of ten graphene layers merged together. That is, the carbon atoms of the sheets experience both the inlayer and interlayer interactions, see  Fig.~\ref{sys1}. Note, that since each carbon atom participates in four covalent bonds, the interaction between the surface atoms and the aluminum atoms is relatively weak. To prevent the vertical motion of the stack of the layers, subject to the normal load, we fixed the bottom layer (shown yellow in Fig.~\ref{sys2}). Moreover, to prevent the horizontal motion of the layers, subject to tangential stress we also fixed the side edges of the graphene layers. 

 For the potential, we use charge-optimized many-body (COMB3) potential. COMB refers to the second generation of the potential \cite{shan2010charge}, and COMB3 corresponds to the third-generation of COMB potential \cite{liang2013classical}. The four terms of the potential depend on two variables -- the charge $q$ and inter-site distance  $r$: 
 
 \begin{equation}
U [{q}, {r}] =  U^{es} [{q}, {r}]  + U^{short} [{q}, {r}]  + U^{vdW} [{r}] + U^{corr} [{r}],  
\end{equation}
where $U^{es} $ is electrostatic term,  $U^{short}$ is charge-dependent short-range interaction, $U^{vdW}$ is van der Waals interaction, and $ U^{corr}$ is correction term.\footnote{For more information see the comb and comb3 pair\_styles in the LAMMPS documents \cite{combpairstyle}} The equilibrium charge on each atom is calculated by the electronegativity equalization (QEq) method \cite{rick1994dynamical}. This potential is of high accuracy although computationally time consuming.  
 
The COMB3 potential has been utilized to model the interface of graphene and metals like C-Cu, and C-Al. In the particular case of Al-C, we can refer to the reference \cite{zhang2019dynamics} for the graphene-Al interface. In this potential, interactions between Al and graphene are mostly weekly bonded. That is, the interface between Al and graphene remains physisorbing. Moreover, we performed our simulation for $T=300 K$, which implies the lack of aluminum carbide.\footnote{Aluminum carbide formation might happen under vacancy defects in graphene and high enough temperatures. However for the temperatures range of $300-900 K$, without any defects in graphene, the simulations remain without the formation of aluminum carbide \cite{zhang2019dynamics}.} For simplicity, we set here zero charges for Al and C atoms; we expect that this would not have any noticeable impact on the surface fluctuations, which hinder the tangential sliding. 

In our simulations we use  the Langevin thermostat along with NVE time integration scheme (Langevin + NVE). This choice has been motivated by the fact that utilizing the Langevin thermostat supplemented by the NVE integration, results in the  simulations of a canonical ensemble with nearly conserved energy \cite{schneider1978molecular}.\footnote{It is important to mention that during thermostating, even in the Langevin dynamics, the removal of the additional energy from the system is not quite steady. Therefore, an unphysical dissipation might be introduced in the system \cite{vanossi2013colloquium}.\label{footthermo}}  We believe that this kind of setup would be the best one for an adequate modelling of  the surface thermal fluctuations. It should be noted however, that the properties of surface fluctuations would depend on the damping parameter of the  Langevin thermostat, which is set with some uncertainty. Indeed, the damping parameter is related to the viscous properties of the solid material -- its bulk dissipative constants, and hence may be defined from these quantities. Unfortunately, neither the bulk dissipative constants, for the material built up of graphene sheets, nor the microscopic theory for such a relation are currently available. Therefore,  we use here plausible values for the damping parameter of the Langevin thermostat, which implies that the simulation results provide qualitative behavior of the system. During the current simulations we used the damping parameter $damp = 0.1$.\footnote{For more information see the fix Langevin command in the LAMMPS documents \cite{langevincom}}

The simulations begin with the application of the normal load $F_{\rm N}=F_z$ distributed over all aluminium atoms of the sphere, Fig. \ref{sys2}. In addition the particle gets an acceleration subject to the applied tangential force $F_x$. The equations of motion for all particles are solved by the Verlet algorithm  \cite{frenkel2001understanding} for a constant temperature of $T=300 K$. After some time, the acceleration is expected to cease and the particle
reaches a constant velocity regime. Here the friction force is equal to the tangential force, $F_{\rm fr} =F_x$. In practice, the relaxation time towards the steady state may be rather long, therefore we apply some tricks to estimate the steady-state sliding velocity using relatively short runs; these are discussed below. 

In applying different normal and tangential forces we observe that efficient and adequate simulations require some limitation for the range of the exerted forces $ F_x$ and $F_z$. To remain in a realistic simulation time, around
a few nanoseconds, we have to choose $F_x$ not too large (to reduce the relaxation time to the steady state) and $F_z$ not too small (to increase the friction force, which again results in the decreasing relaxation time). Nevertheless, $F_z$ cannot be too large, as a very large force would mash the particle. All in all, we have chosen the intervals,  $F_x \in [2,5] nN$ and $F_z \in [15, 60] nN$. 

We  simulate a particle moving in the  positive $x$-direction, where the  periodic boundary condition have been applied. This allows  long-time simulations. We use the simulation time step of the order of femtosecond ($dt \sim 0.001 \times 10^{-12} s$), and the total simulation time was about a few nanoseconds ($ \sim  10^{-9} s$).

\subsection{Estimation of the steady state velocity} 

We conjecture that the friction force is proportional to  the sliding velocity, $F_{\rm fr}=-\gamma v$.  Consider the equation  of motion for a particle moving on the surface,  subject to the tangential force $F_x$, friction force $-\gamma v$ and a random force $\zeta(t)$ with zero mean, $\left< \zeta(t) \right>=0$, 
$$
m \dot{v}=-\gamma v +F_x +\zeta(t). 
$$
After averaging the above equation we obtain the time evolution of the average velocity, when the initial average velocity is zero, $ \langle v (0)\rangle =0$:
\begin{equation}
 \langle v (t) \rangle = \frac{F_x}{\gamma} \left[ 1 - e^{- (\gamma/m) t} \right], 
\label{Eqvelo}
\end{equation}
which describes the relaxation of the average velocity to its steady value 
$$
\left. \langle v \rangle\right|_{t \to \infty} = v_{st} = \frac{F_x}{\gamma}.
$$
The last equation shows that the friction coefficient may be found from the (known) tangential force $F_x$ and the steady velocity. 

Hence, to measure $\gamma$ in computer simulations, one needs to wait until the steady state is achieved. In reality,  however this is not very practical, since the relaxation time may be rather long. There is a couple of ways to shorten the simulation time. Firstly, one can use  Eq. \eqref{Eqvelo} and apply a fitting procedure,  to find the best fit  for $\gamma$ to satisfy this relation. Secondly, one can utilize a more simple approach, which we use here. Suppose that we require the accuracy of $\alpha$ (say $\alpha \approx 0.99$) for the estimate of $\gamma$. Then at the first step we roughly estimate $\gamma$ and at the second step we predict the duration of the simulation time $t_s$, such that the average velocity $\langle v (t_s) \rangle$    at $t=t_s$ approximates the steady state velocity $v_{st}$, that is, $\langle v (t_s) \rangle \approx v_{st}$,  with the accuracy   $\alpha$.  It immediately follows from Eq. \eqref{Eqvelo} that  
\begin{equation}
t_s \simeq  - \frac{m}{\gamma} \, \ln (1 - \alpha ) . 
\label{ptn}
\end{equation}
In our MD experiments  we use the above estimate for the total simulation time.

\section{Results and discussion} 
Fig.~\ref{sys1:first} depicts the system at the onset of the simulation, while Fig.~\ref{sys1:second}, after some simulation time, when the  surface thermal fluctuations have developed. During the simulations, the temperature was kept around $300 K$.

\begin{figure}[h]
\centering
\includegraphics[width=0.99\textwidth]{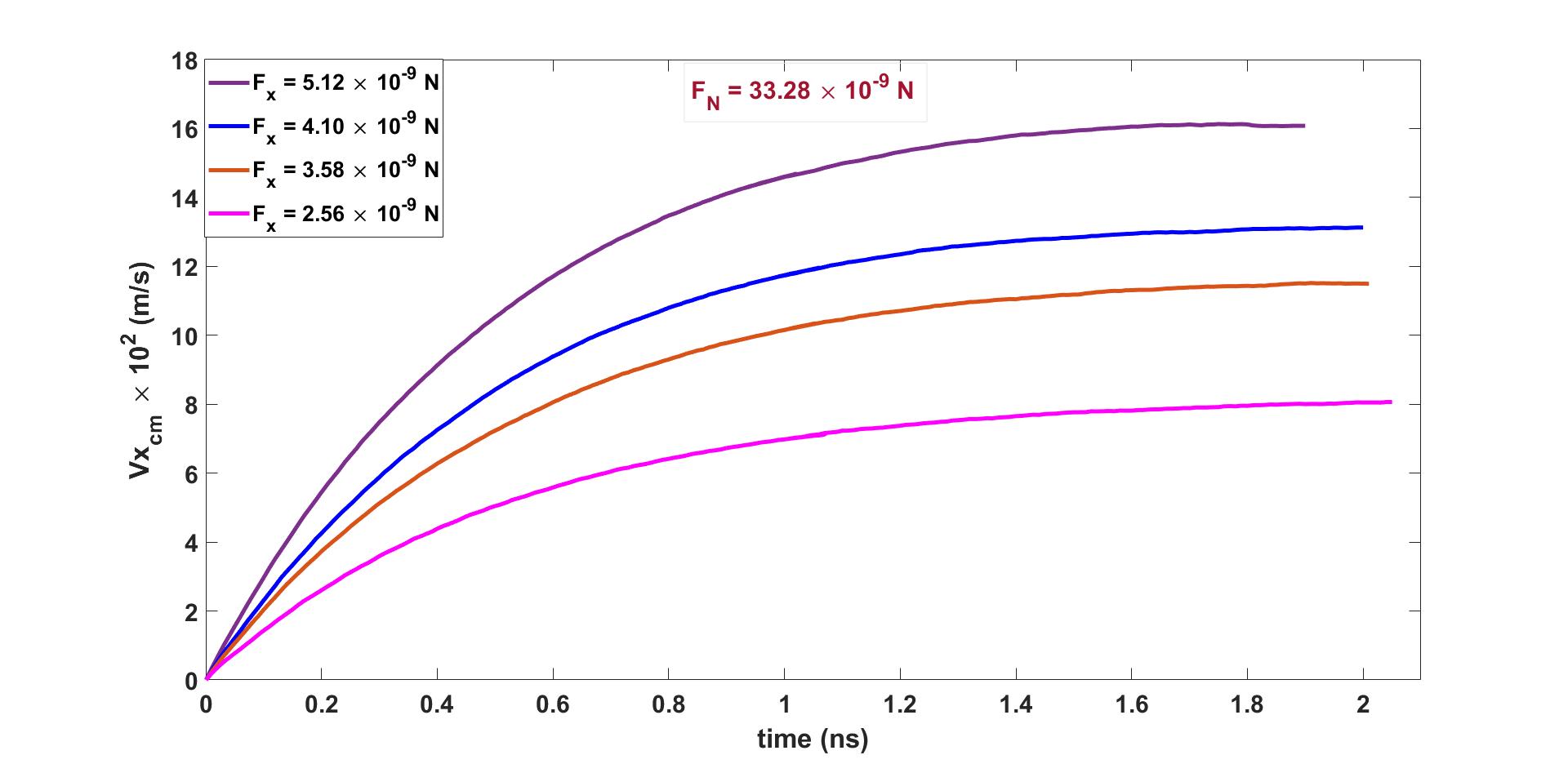}
\caption{Simulated steady-states for the normal load of $F_z = 33.28 \times 10^{-9} N$, and four different horizontal forces $F_x$.}
\label{stwrdy}
\end{figure}

\begin{figure}
\centering
\includegraphics[width=.99\textwidth]{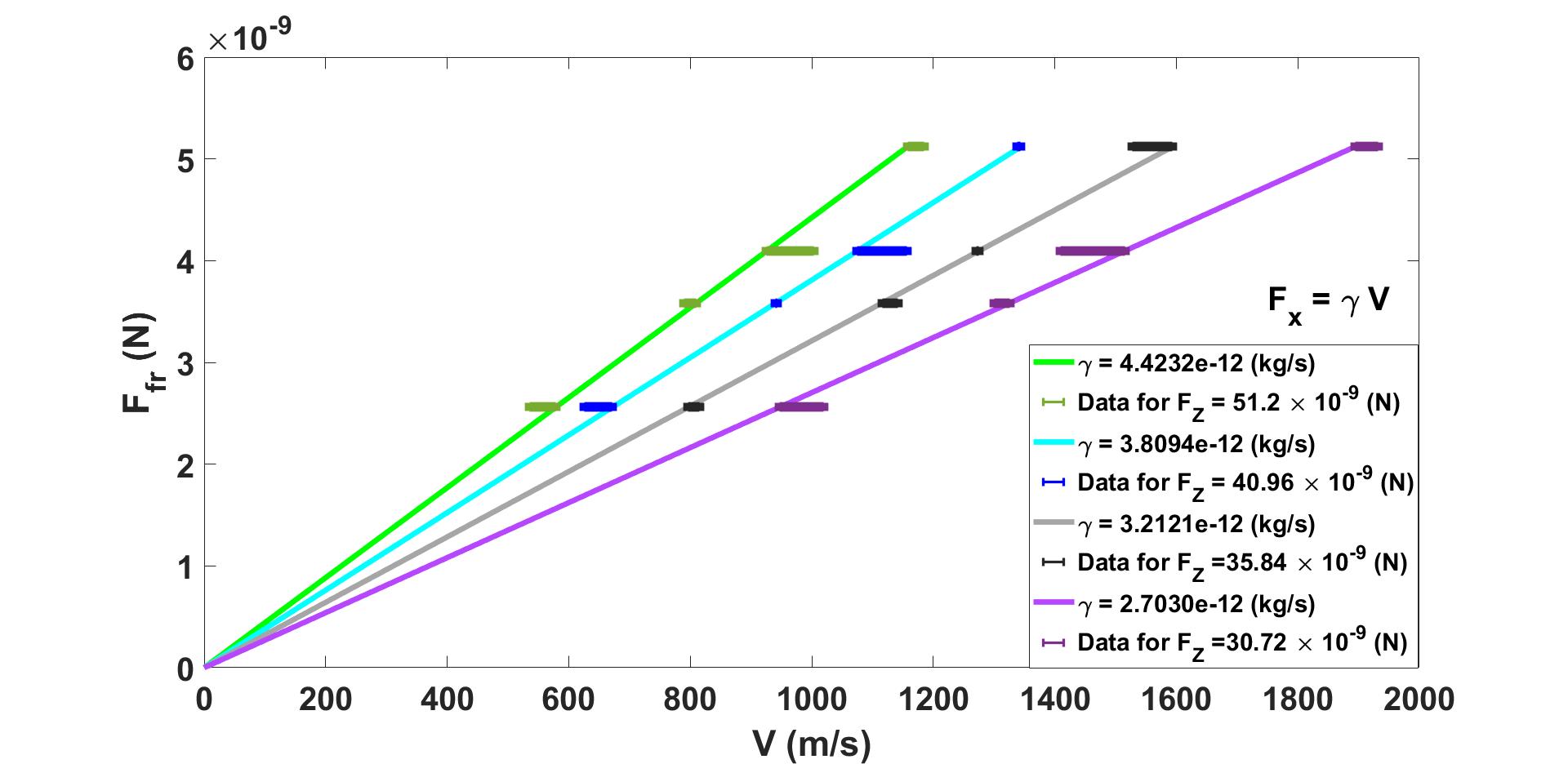}
\caption{Illustration of the linear dependence of friction force on the steady state sliding velocity for different normal load.  }
\label{4figures}
\end{figure}

Fig.~\ref{stwrdy} illustrates four realizations of the velocity relaxation to a nearly steady state regime for a constant normal load of $F_z = 33.28 \times 10^{-9} N$  and different horizontal forces $F_x$. As it may be seen from the figure, the total simulation time was about $2ns$, which suffices to obtain good estimates for the friction coefficient. 
Fig.~\ref{4figures} demonstrates the linear dependence of the  friction force on the velocity of the nanoparticle in the steady state sliding. The best fit yields,  $F_{x}= -\gamma v_{st}$, where $v_{st}$ has been obtained as discussed above. 

Now we shortly discuss the possible errors in our simulations. As it has been stated above, one source of errors is the lack of true steady states, and the approximation of the steady state  velocity by its approximate transient value. The fixation of the edges of the graphene layers violates the symmetry of surface fluctuations with respect to $x$ and $y$ directions. This may result in some unphysical effects with an impact on the friction force. Finally, the Langevin thermostat, as a stochastic setup also contributes to the errors. In the latter case, however, the error may be mitigated by multiple runs. To check the  reliability  of the measured friction coefficient $\gamma$ we perform a couple of independent runs with the same system parameters. We observe that  the estimate is reliable for the most of the normal forces $F_z$, except for the smallest loads.  Here the  linearity of friction force on sliding  velocity could be questionable.

\section{Conclusion} 

We investigate numerically, by the means of MD simulation, dry sliding friction at the nanoscale. In our study we utilize LAMMPS with the most modern  COMB3 many-body potentials for the inter-atomic interactions. In our simulation setup we use a spherical  particles comprised of 32,000 aluminum atoms, which rests on a semi-infinite space with a free surface. The latter part  of the system is modelled by a stack of merged graphene layers, with a fixed bottom layer. We use the Langevin thermostat which keep the  temperature  at $300 K$. We vary the normal load on the particle $F_z$ and apply different tangential force $F_x$. We observe that after a transient time, the system relaxes to a state with a steady sliding velocity $v_{st}$. The steady velocity is determined by the applied tangential force or, the other way around, the friction force is determined by the sliding velocity. Hence, based on our data of the MD experiments we demonstrate that the friction force linearly depends on the sliding velocity, that is, $F_{\rm fr}= -\gamma v$. Such a linear dependence holds true for the studied interval of normal and tangential forces, $ F_z \in  [15,60]  \,\, nN$ and  $F_x  \in  [2,5]  \,\, nN$. It, however differs from the non-linear dependence expected from the Prandtl-Tomlinson (PT)model \cite{Vanossi2013}. Although the agreement of our simulation data with the predictions of the  PT model  was not expected, the relatively large sliding velocities used in our study may be among the reasons of the difference. 

The main conclusion of our study is that, in contrast to the macroscopic dry sliding friction, the friction force at the nanoscale is velocity-dependent. Moreover for the range of normal loads and sliding velocities explored here, the friction force linearly increases with the increasing velocity. The physical mechanism of the observed friction behavior is the corrugation of the contact plane by surface fluctuations of the  thermal origin. Such fluctuations hinder the relative motion of the surfaces, yielding  the friction force, proportional to the sliding velocity. This is similar to the viscous friction force that acts on a solid body moving in  viscous fluid.  
\ack

AAT and NVB gratefully acknowledge the Russian Foundation for Basic Research (RFBR), Grant No. 18-29-19198. The research was carried out using the equipment of the shared research facilities of HPC computing resources at Lomonosov Moscow State University \cite{adinets2012job} and Skoltech supercomputer Zhores \cite{zacharov2019zhores}.

\section*{References}
\bibliography{dry_fric_V}

\end{document}